  \documentclass[twocolumn,prb,aps]{revtex4}

 \usepackage[dvips]{graphicx}
 \usepackage[dvips]{graphics}

\begin{document}
\title{Absence of an isotope effect in the magnetic resonance in high-$T_c$ superconductors}
\author{S. ~Pailh\`es$^1$, P.~Bourges$^{1\ast}$, Y. ~Sidis$^1$, 
C. Bernhard$^{2}$, B. Keimer$^{2}$, C.T. Lin$^{2}$, J.L. Tallon$^{3}$}

\affiliation{
$^1$ Laboratoire L\'eon Brillouin, CEA-CNRS, CEA-Saclay, 91191 Gif sur Yvette, France.\\
$^2$ Max-Planck-Institut f\"ur Festk\"orperforschung, 70569 Stuttgart, Germany\\
$^3$ Industrial Research Limited and Victoria University, P.O. 31310, Lower Hutt, New Zealand.\\
}


\begin{abstract}

An inelastic neutron scattering experiment has been performed in the high-temperature 
superconductor $\rm YBa_2Cu_3O_{6.89}$ to search for an oxygen-isotope shift of the 
well-known magnetic resonance mode at 41 meV. Contrary to a recent prediction 
(I. Eremin, {\it et al.}, Phys. Rev. B {\bf 69}, 094517  (2004)), a negligible shift 
(at best $\leq$ +0.2 meV) of the resonance energy is observed upon oxygen isotope 
substitution ($^{16}$O$\rightarrow^{18}$O). This suggests a negligible spin-phonon interaction in the high-$T_c$ cuprates at optimal doping.
\end{abstract}

\maketitle

In conventional superconductors, pairing between electrons is mediated by lattice 
vibrations\cite{BCS}. This has been demonstrated by an isotope effect 
on the superconducting (SC) transition temperature, $T_c$. 
In high-$T_c$ copper oxides superconductors $T_c$
exhibits a weak shift at optimal doping upon isotope substitution \cite{franck} 
which increases at lower doping.
In particular, the oxygen-isotope shift ($^{16}$O$\rightarrow^{18}$O) has been 
extensively studied\cite{zhao,pringle,williams}. 
At optimal doping, a small isotope-effect exponent
is deduced $\alpha_{T_c}= - d \ln T_c /d \ln M  \simeq 0.05$ 
much lower than the $1\over2$ value expected from pure electron-phonon interaction, casting some doubt on a superconducting mechanism mediated by phonons. Further, the proximity of the antiferromagnetic (AF)
insulating state and the unconventional $d$-wave symmetry of the SC gap favored 
mechanisms for high T$_c$ superconductivity where electron-electron (el-el) interactions 
predominate. However, there has been a revival of interest in electron-phonon coupling as several experiments point towards a non-negligible electron-phonon 
interaction\cite{lanzara,keller,gweon}. In particular, the ``kink'' change of slope of electronic dispersion observed around $\sim$ 70 meV by 
angle-resolved photoemission spectroscopy (ARPES) along the nodal direction in various cuprates  could be interpreted 
as an electronic coupling to a phonon mode \cite{lanzara}. As a matter of fact, 
various physical properties such as penetration depth\cite{keller} or ARPES spectra
 \cite{gweon} display relatively large isotope effects thus highlighting the open question: 
what can be the role of phonons in determining the superconducting properties of
cuprates?

On the other hand, the spin excitation spectrum of the copper oxide superconductors
is particularly rich. Above $T_c$, magnetic fluctuations are mainly observed around 
the planar wavevector ${\bf Q_{AF}}\equiv \rm (\pi/a,\pi/a)$ characteristic 
of antiferromagnetism (AF) in the undoped parent compounds 
\cite{rossat,bourges,fong00,dai01}. Below $T_c$, a collective 
magnetic mode, referred to as the ``resonance peak'', appears at a well-defined 
energy\cite{rossat,fong95,bourges,fong00,dai01,Sidis03} at $(\pi/a,\pi/a)$ and 
exhibits strong dispersions for wavevectors around ${\bf Q_{AF}}$\cite{Bourges-science,Pailhes04,Reznik04}.
This mode is now observed in all high-$T_c$ superconductors systems studied by inelastic 
neutron scattering (INS) experiments \cite{rossat,fong99,he02} whose maximum 
$T_c$ reaches 90 K. Recently, an analogous feature has even been reported in the 
single-layer material $\rm La_{2-x}Sr_x Cu O_{4}$ as well\cite{jmt04}. 
Depending, or not, whether the magnetic fluctuations are observed in the normal state, 
the resonance peak either corresponds to a modification of magnetic spectrum in both momentum
and energy (in underdoped cuprates) \cite{bourges,fong00,dai01} or simply emerges from 
the magnetic electron-hole continuum (in optimally doped and overdoped cuprates) 
\cite{Sidis03,Reznik04,Pailhes03}. 

This mode is typically assigned to an excitonic bound state in the superconductivity-induced 
gap in the spectrum of electron-hole spin-flip Stoner excitations
\cite{vdM,Liu95,Millis96,chubukov,Onufrieva02,norman,abanov,Eremin,Eremin04}. Within that 
framework, the resonant mode is a direct consequence of unconventional superconductivity 
of $d$-wave symmetry occurring in the high-$T_c$ copper oxides.  
This approach is particularly suited for optimally doped superconductors where normal-state magnetic fluctuations are consistent with a broad magnetic electron-hole continuum.
Further, the most recent neutron developments \cite{Pailhes04} allow "silent bands" (where the magnetic collective mode is overdamped) to be related to the 
detailed momentum shape of the Stoner continuum expected from the Fermi surface topology and the $d$-wave superconducting order parameter. This connection has been explicitly assigned in ref. \cite{Eremin04}. This spin exciton mode can be derived from an effective $t-J$ Hamiltonian 
\cite{Onufrieva02,Eremin}. Within that model, an interesting proposal has been 
made \cite{Eremin} that magnetic properties could display significant isotopic 
effects if both the hopping integral, $t$, and the superexchange interaction 
between neighboring spins, $J$, are renormalized by phonons. 
By changing the oxygen isotope $^{16}$O by $^{18}$O, they predict a change of the 
resonance peak position of a few meV, mostly due to a re-normalization of the 
hopping integral. 

Thus, by measuring the isotope dependence of the position of 
the resonance peak, INS could provide direct evidence for the presence of the 
electron-phonon coupling in cuprates. In contrast, we here report the probable absence 
of an isotope effect in the magnetic resonance peak in $\rm YBa_2Cu_3O_{6.89}$ (YBCO).  

The inelastic neutron experiment was performed on the 1T spectrometer
at Laboratoire L\'eon Brillouin (LLB) in Saclay. The spectrometer used a vertically and horizontally focusing monochromator and analyzer, comprising Cu (111) crystals and 
pyrolytic graphite (PG002) crystals, respectively. The measurements were performed
with a fixed final neutron energy of 30.5 meV. A filter was inserted
into the scattered beam in order to eliminate higher order
contamination. The crystals were oriented such that momentum
transfers $\bf Q$ of the form ${\bf Q}\rm =(H,H,L)$ were accessible. We use
a notation in which $\bf Q$ is indexed in units of the tetragonal
reciprocal lattice vectors $2\pi/a=1.63 $\AA$^{-1}$ and
$2\pi/c=0.54 $\AA$^{-1}$. 

\begin{figure}[t]
 \includegraphics[width=7.5 cm,height=6cm]{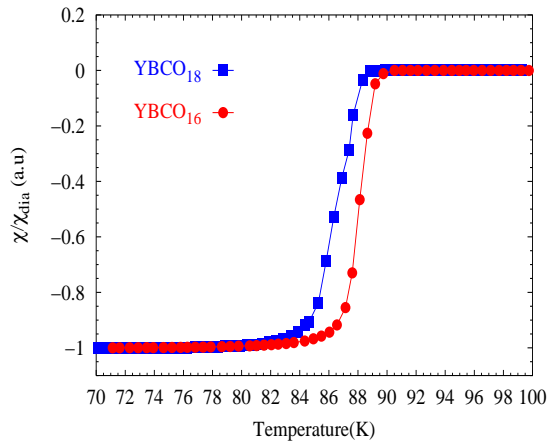}
\caption {
{\label{fig1}} (Color online)  Magnetic susceptibility of both sample arrays used in this experiment. 
Onset superconducting temperatures occur at 88.8 K for samples with $^{18}$O and 90 K 
for $^{16}$O. The curve of each mounting has been obtained by weighting the 
susceptibility curve of each individual single crystal by their mass. The dispersion 
of $T_c$ for the samples of each array is less than 0.3 K. }
\end{figure}

High-quality single crystals of $\rm YBa_2 Cu_3 ^{16}O_{7-\delta}$, of typical mass 
$\sim$ 0.1-0.2 g, were prepared. They have been separated in two distinct sets of similar
total mass ($\sim$ 0.6 g). The two batches of samples were mounted in a furnace in 
separate quartz tubes side by side to ensure identical thermal history. One tube was 
charged with high purity $^{16}$O oxygen gas while the other was charged with 
99\% $^{18}$O enriched oxygen. After annealing for 24 hours at 830 $^\circ$C the pair 
of tubes was removed, evacuated, recharged and reannealed for a total of 10 exchanges 
to ensure maximal isotope exchange. Both sample sets were then slow cooled over 48 hours 
to 550 $^\circ$C then annealed there for 10 days to ensure the same uniform oxygenation 
of the chains. The samples were lightly underdoped as confirmed by thermoelectric power 
measurements\cite{Obertelli}. Based on previous annealing experience for YBCO ceramics 
and crystals\cite{jeff} we expect the samples to have an oxygen content of 
$\approx$ 6.89 and an underdoped hole concentration of $n_h \approx$ 0.145. 
The magnetic susceptibility for each individual sample has been measured by a  SQUID (
superconducting quantum interference device) magnetometer. 
The crystals for each individual sample were then aligned on two distinct arrays
(referred hereafter as S$_{16}$ and S$_{18}$) of
similar volume, each array being made of about four single crystals. The magnetic
susceptibility measurements of each sample mounting are shown on Fig. \ref{fig1}: 
one can identify the onset of the superconducting transition at 90 K for S$_{16}$ and 
88.8 K for S$_{18}$. A difference in $T_c$ of $\sim$ -1.2 K is then observed between both 
samples. It is actually larger than the reduction expected from the usual isotope effect which 
is about -0.77 $\pm$ 0.2 K for similarly-doped YBCO \cite{pringle}. The slight excess in 
isotopic shift in $T_c$ is possibly significant and may suggest a slightly lower oxygen 
content and doping state for the $^{18}$O sample, despite the attempt to ensure identical 
thermal histories. To address this issue, we further determined the $c$-lattice parameter 
of both samples. Accurate measurements using the triple axis 4F1 spectrometer installed 
on a cold source at LLB yield $c$= 11.674 \AA $\pm$  0.004 for S$_{16}$ and 
$c$= 11.678 \AA $\pm$ 0.004 for S$_{18}$. 
These $c$-lattice parameters are consistent with an oxygen content of about 
$x$=0.89\cite{jorgensen,cava}, and given the slope $\partial c/\partial x = -0.11$ 
\AA\cite{jorgensen,cava}, indicates that the $^{18}$O sample has a possible lower 
oxygen content by $\Delta x = 0.02 \pm 0.03$. Thus, it is in agreement with the 
possibility that part of the difference in $T_c$ for each sample is related to a 
slight difference in doping. Using the relationship between oxygen content and the 
hole doping\cite{jeff}, this would correspond to a difference of 
$\sim$ $\delta n_h$=0.004 $\pm$ 0.006 in doping. 

\begin{figure}[t]
 \includegraphics[width=7.5 cm,height=7cm]{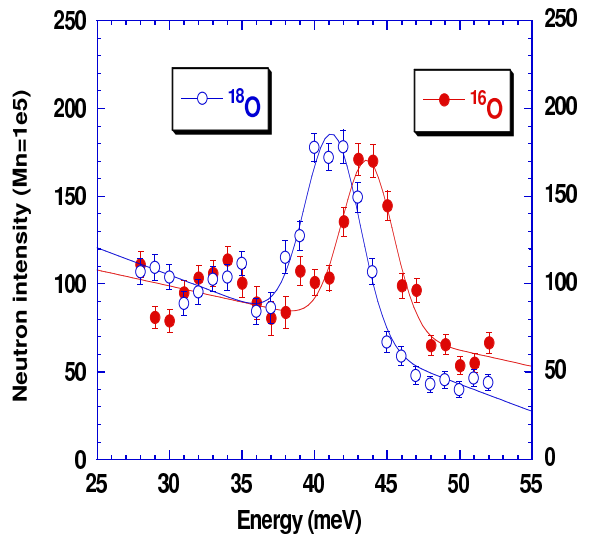}
\caption {
{\label{fig3}} (Color online)  Neutron intensity measured at Q=(-0.5,-0.5,10.3) and at T=100 K showing 
an oxygen phonon mode in both $^{16}$O sample and in the $^{18}$O sample. }
\end{figure}

In order to check the isotope exchange process, we performed Raman scattering 
at room temperature as well as INS measurements of a particular oxygen phonon mode. 
The three oxygen modes observed in Raman scattering, namely the $c$-axis vibration of 
the apical oxygen and the in-phase and out-of-phase oxygen vibrations in the CuO$_2$ 
plane, revealed isotopic shifts to lower energy in $^{18}$O  sample with respect to 
the $^{16}$O sample. Assuming that they are pure oxygen modes, the fraction of 
exchanged oxygen is y$\simeq 0.95 \pm 0.05$. 
Being a surface-sensitive technique, Raman scattering does not indicate if
the isotope exchange occurred within the bulk of the samples. This might 
be problematic as oxygen diffusion is very slow. 
We then measured with inelastic neutron scattering a particular phonon mode
whose eigenvector is predominantly related to a vibration along the c axis
of the oxygens of the CuO$_2$ plane\cite{fong95}. This phonon mode, 
measured at ${\bf Q}$=(-0.5,-0.5,10.3) where its structure factor is larger, is found 
at 43.6 $\pm$ 0.1 meV in S$_{16}$ and 41.3 $\pm$ 0.1 meV in S$_{18}$ 
(Fig. \ref{fig3}). Again assuming that the phonon is a pure oxygen mode,
this corresponds to an isotope exchange of $y$=0.92 $\pm$ 0.08. In fact, the eigenvector 
for that specific mode corresponds to about 90 \% weighting by the oxygen atoms.
Therefore, the energy shift of the phonon measured in INS is fully consistent 
with the nearly full oxygen exchange deduced from the Raman data. 
The oxygen exchange thus occurred throughout the bulk of the material thus confirming 
the sample homogeneity. 

\begin{figure}[t]
\includegraphics[width=7.5 cm,height=8 cm]{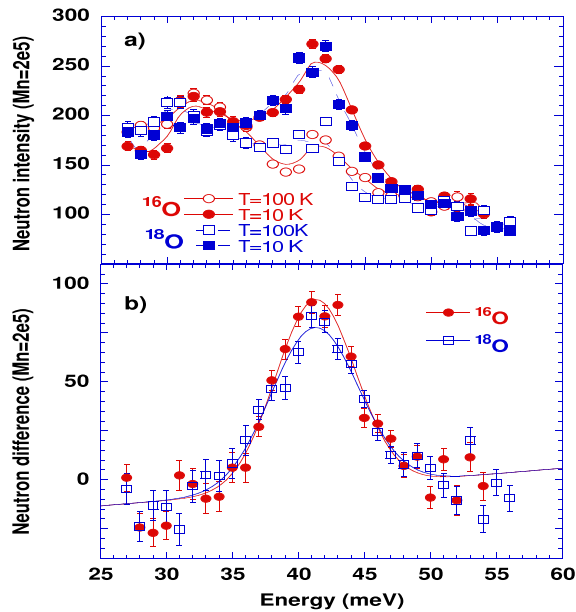}
\caption {
{\label{fig4}} (Color online) Resonant magnetic peak intensity in both $^{16}$O samples 
and $^{18}$O samples: a) Raw neutron intensity at 12 K and 100 K for both samples, 
b) Difference of the neutron intensity 12 K - 100 K for each sample. The total 
counting time reached 1.5 hour per point at each temperature to reduce the error 
bar on the energy of the resonance to 0.12 meV. The curves are not normalized by 
samples mass. }
\end{figure}

We now turn to the central result of this report. Following previous studies
\cite{fong95,bourges,Pailhes04,Reznik04}, we identify the resonant magnetic mode 
by constructing, for each sample, the difference between constant-$\bf Q$ scans 
measured at 12 K  ($\rm <T_{c}$) and 100 K ($\rm >T_{c}$) and at the wave vector 
${\bf Q}$=(-0.5,-0.5,5.1). The magnetic resonant mode in both samples is shown 
in Fig. \ref{fig4}.b  as well as the raw energy scans for both samples in 
Fig. \ref{fig4}.a. At T=100 K, the background, which displays the same phonon 
mode as the mode shown in Fig. \ref{fig3} but with much weaker intensity\cite{fong95}, 
is also shifted upon the isotope substitution. This phonon is known to exhibit no
temperature dependence across $T_C$. Further, the phonon scattering structure 
factor for the chosen wavevector with $L$=5.1 is reduced enough that the background 
subtraction procedure in the determination of the resonance peak energy, 
its broadening and its amplitude does not affect significantly the observed 
difference of  Fig. \ref{fig4}.b. The resonance peak energy is found at 
${\omega_R}=41.26 \pm$ 0.12 meV in S$_{16}$ and 
${\omega_R}=41.31 \pm$ 0.12 meV in S$_{18}$. The two energies are therefore not 
distinguishable within errors. The two peaks exhibit a slight difference in
amplitude as well as in width (7.1 $\pm$ 0.3 meV for S$_{16}$ and and 7.8 $\pm$ 
0.3 meV for S$_{18}$). In principle, the observed width is not intrinsic but 
is controlled by the convolution product of dispersive excitations around 
$(\pi,\pi)$ with the spectrometer resolution \cite{Bourges-science,Pailhes04}.
The product of the peak amplitude by its width in Fig. \ref{fig4}.b, representing 
the magnetic resonant spectral weight at $(\pi,\pi)$, is similar in both
samples within errors. Using the spectral weight of the phonon presented in 
Fig. \ref{fig3}, one can calibrate the absolute magnetic intensity of 
the resonance peak\cite{fong95,fong00}. For both samples, we deduce an energy-integrated 
magnetic spectral weight of 2.6$\pm$0.4  $\mu_B^2$ at the $(\pi,\pi)$ 
wavevector, or 0.06$\pm$0.01 $\mu_B^2$ for energy- and q-integrated magnetic 
spectral weight in agreement with 
a previous report for a similar doping level\cite{fong00}. 

We then basically observe no isotope effect of the magnetic resonance 
peak: $\alpha_{\omega_R}= - d \ln {\omega_R} /d \ln M \simeq 0$. 
To be complete, there is however the possibility of a slight difference in doping between 
the two samples which might induce a slight change in the resonance energy if the 
resonance energy is proportional to $T_c$, as it is typically observed 
\cite{bourges,fong00}. (To what accuracy this proportionality  strictly applies is 
still an open question). According to this empiric relation, the resonance peak energy 
could be renormalized by about 0.5\% in the $^{18}$O sample as compared with the 
$^{16}$O sample, i.e. an energy shift of $\sim$ -0.18 meV. In such a case, an isotope 
effect can be estimated of $\delta{\omega_R} \simeq$ +0.23, yielding an isotope exponent 
of $\alpha_{res}= -0.05$. Therefore, the isotope shift of the resonance peak energy 
can be {\it at most} $\delta {\omega_R} \leq$ 0.23 meV $\pm$ 0.2 meV. The deduced 
isotope-effect on the resonance peak is then very small and actually similar in 
magnitude to the small isotope effect of the superconducting  transition for optimally 
doped cuprates, $\alpha_{T_c}= 0.05$\cite{franck}, although with an opposite sign. 
It should be noticed that the overall effect might be simply overshoot by the difference in doping between both samples. 

The absence of an isotope effect on the resonance energy is actually quite surprising as, in the spin exciton model, the bound state energy is very sensitive to both band structure, via the hopping integrals $t$,$t'$..., and the interactions, $g$. Within a random-phase approximation (RPA) scheme, 
the resonance energy at the AF wavevector is usually defined  as the pole of the interacting susceptibility, $1- g/2 Re \chi_0(Q_{AF},\omega_R)=0$ 
\cite{Liu95,Millis96,chubukov,Onufrieva02,norman,abanov,Eremin,Eremin04}, 
where $\chi_0(Q,\omega) \propto 1/t$ is the bare spin susceptibility of a $d$-wave superconductor. The interaction $g$ can be either the superexchange interactions\cite{Liu95,Millis96,norman,Onufrieva02,Eremin}, 
$4 J_{AF}$ at $(\pi,\pi)$, or some spin-fermion coupling \cite{abanov,chubukov,Eremin04}.  
Looking in more detail at the pole condition, one clearly sees that a shift in $g$ and in $t$ have opposite effects on the resonance energy $\omega_R$: if the interaction is reduced, the resonance energy will increase whereas if the band hopping integral is reduced the resonance energy will decrease. Using such a model, Eremin et al\cite{Eremin} expected a shift of about -2 meV of the resonance peak from $^{16}$O to $^{18}$O, corresponding to $\alpha_{\omega_R}= 0.4$. That was because the band structure hopping integral $t$ is thought to 
be the quantity strongly dependent on the electron-phonon coupling. 
As the observed $\delta_{\omega_R}$ has an opposite sign (if any) of the theoretical prediction \cite{Eremin}, it is doubtful that the electron-phonon coupling is renormalizing the band structure hopping integral. Within that model, one can nevertheless explain the observed positive sign of 
$\delta_{\omega_R}$  by a small renormalization of the interaction $g$ rather than $t$.
Our maximum estimate of $\delta_{\omega_R} \le 0.23$ meV would typically yield
$\delta g/g \leq$ - 0.4 \%, in agreement with the expected impact of the interactions 
term\cite{Eremin}. More specifically,  if the interaction is assigned to the AF 
superexchange, we obtained an isotopic change of $J_{AF}$ compatible with the one, 
$\delta J / J \sim$ - 0.6 \% \cite{zhao}, deduced from the N\'eel temperature in 
undoped cuprates. Finally, this simple analysis of the tiny shift (if any) of the 
resonance peak energy in term of the 
spin-exciton model shows that the various microscopic terms entering in its
expression do not exhibit a strong isotope effect, thus placing a severe limit on 
the role of electron-phonon coupling in high-$T_c$ cuprates.

In conclusion, using inelastic neutron scattering experiments, we observe no significant 
shift of the magnetic resonance peak energy in YBCO upon substitution of oxygen 
$^{16}$O by its isotope $^{18}$O. In contrast to previous claims, this suggests that 
the spin-phonon coupling is negligible in high-$T_c$ cuprates near optimal doping. 
The absence of a measurable effect on the INS resonance mode does not however exclude 
the possibility that isotope substitution can have a noticeable effect on the magnetic 
properties at much lower doping. 

We thank Ilya Eremin and M.V. Eremin for stimulating discussions, alerting us to possible isotope effects in the resonance energy.

\end{document}